\documentclass[12pt]{article}

\usepackage{amsmath,amssymb,cite}
\usepackage{graphicx,color}

\newcommand{\f}[2]{\frac{#1}{#2}}
\newcommand{\ko}[1]{\left( #1 \right)}
\newcommand{\kko}[1]{\left[ #1 \right]}

\newcommand{\abs}[1]{\left| #1 \right|}
\newcommand{\ket}[1]{\left| #1 \right\rangle}
\newcommand{\kket}[1]{| #1 \rangle}
\newcommand{\bra}[1]{\left\langle #1 \right|}
\newcommand{\kbra}[1]{\langle #1 |}
\newcommand{\bmt}[1]{{{\mbox{\boldmath$ #1 $}}}}

\newcommand{\ip}[2]{\langle #1 | #2 \rangle}

\DeclareMathOperator{\diag}{diag}
\DeclareMathOperator{\tr}{tr}

\def\ds{\displaystyle}

\def\eq{\equiv}
\def\be{\beta}

\def\vp{\varphi}
\def\no{\nonumber}
\def\lam{\lambda}
\def\ep{{\epsilon}}
\def\dag{{\dagger}}

\def\N{{\mathcal N}}

\def\cW{{\mathcal W}}
\def\cR{{\mathcal{R}}}
\def\eK{\mathrm{\bf K}}
\def\eE{\mathrm{\bf E}}

\numberwithin{equation}{section}
\begin{document}

\quad 
\vspace{-3.0cm}

\begin{flushright}
{\bf December 2006}\\
UT-06-24 \\
{\tt hep-th/0612269}
\end{flushright}

\vspace*{0.5cm}

\begin{center}
\Large\bf 
Emergent Classical Strings from Matrix Model
\end{center}
\vspace*{0.7cm}
\centerline{
Yasuyuki Hatsuda${}^{\ast}$ ~~and~~ Keisuke Okamura${}^{\dagger}$
}
\begin{center}
\emph{Department of Physics, Faculty of Science, University of Tokyo,\\
Bunkyo-ku, Tokyo 113-0033, Japan.} \\
\vspace*{-0.5cm}
\begin{align}
\mbox{${}^{\ast}$\,{\tt hatsuda}}&\mbox{{\tt @hep-th.phys.s.u-tokyo.ac.jp}}\nonumber \\
\mbox{${}^{\dagger}$\,{\tt okamura}}&\mbox{{\tt @hep-th.phys.s.u-tokyo.ac.jp}}\nonumber
\end{align}
\end{center}

\vspace*{0.7cm}

\centerline{\bf Abstract} 

\vspace*{0.5cm}

Generalizing the idea of {\tt hep-th/0509015} by Berenstein, Correa, and V\'{a}zquez, we study many-magnon states in an SU(2) sector of a reduced matrix quantum mechanics obtained from $\N=4$ SU$(N)$ super Yang-Mills on ${\mathbb R}\times S^{3}$.
Generic $Q$-magnon states are described as a chain of ``string-bits'' joining $Q+1$ eigenvalues of background matrices which form a 1/2 BPS circular droplet in the large $N$ limit.
We will concentrate on infinitely long states whose first and last eigenvalues localize at the edge of the droplet.
Each constituent string-bit has a complex quasi-momentum in general, while the total quasi-momentum $P$ of the state is real.
For given $Q$ and $P$, the minimum energy of the chain of string-bits is realized when the $Q+1$ eigenvalues are equally spaced on one and the same line segment joining the two outmost eigenvalues localized on the edge with angular difference $P$.
Such configuration of bound string-bits precisely reproduces the dispersion relation for dyonic giant magnons in classical string theory.
We also show the emergence of two-spin folded/circular strings in special infinite spin limit as particular configurations of closed chains of string-bits.

\vspace*{1.0cm}

\vfill

\thispagestyle{empty}
\setcounter{page}{0}

\newpage
\section{Introduction}

The correspondence between the type IIB string theory on $AdS_{5}\times S^{5}$ and the four-dimensional $\N=4$ SU($N$) super Yang-Mills theory (SYM) is the best studied example of the AdS/CFT correspondence \cite{Maldacena:1997re}.
Lots of tests have been done to see if it is an exact quantum duality, and if so, why should it be the case. 
So far there has been been no apparent breakdown of the duality, and one can naturally seek for the ways to reconstruct not only the geometries where the string theory is defined but also the massive string modes on them, from what we have in the dual SYM theory.
The first step toward the program was taken in \cite{Berenstein:2005aa,Berenstein:2005jq,Berenstein:2005ek,Berenstein:2006yy}, where new light on the duality was shed from matrix theory point of view, without the apparent need of integrability.\footnote{\,By contrast, in recent years much progress in testing the AdS/CFT has been based on integrable structures of both theories, particularly the ones that can be captured by the Bethe ansatz method.}
In \cite{Lin:2004nb}, Lin, Lunin, and Maldacena (LLM) classified all the 1/2 BPS solutions in the type IIB supergravity in terms of certain boundary conditions or ``droplets'', which confirmed the earlier classifications of corresponding SYM operators by Berenstein \cite{Berenstein:2004kk}.
Recently in \cite{Vazquez:2006id}, a new way to reconstruct the 1/2 BPS metrics from the dilatation operators of SYM side has been proposed.

It is also important to study massive string modes on those geometries.
In \cite{Berenstein:2005jq}, Berenstein, Correa, and V\'{a}zquez (BCV) showed a useful approach to compute the anomalous dimension of Berenstein-Maldacena-Nastase (BMN) states \cite{Berenstein:2003gb} at strong coupling.
They used a gauged matrix quantum mechanics to reproduce the geometry of $S^{5}$ and described the BMN states by using the so-called ``string-bits'', which correspond to off-diagonal modes of the matrices.
Their formalism with a simple saddle point approximation lead to the BMN energy formula to all-order in the 't Hooft coupling $\lam\eq g_{\rm YM}^{2}N$ quite successfully.
Actually they could also imply the existence of the so-called giant magnon of Hofman and Maldacena \cite{Hofman:2006xt}, which is an open string solution with an infinite angular momentum.\footnote{\,For references on (generalizations of) giant magnons, see \cite{Dorey:2006dq,Chen:2006ge,Arutyunov:2006gs,Minahan:2006bd,Spradlin:2006wk,Bobev:2006fg,Kruczenski:2006pk,Okamura:2006zv,Kalousios:2006xy,Hirano:2006ti,Ryang:2006yq,Chen:2006gq,Roiban:2006gs,Chu:2006ae}.
In a recent paper \cite{Maldacena:2006rv}, new interpolating limit of $AdS_{5}\times S^{5}$ was considered. 
It was shown their limit connecting the pp-wave \cite{Berenstein:2003gb} and the giant magnon \cite{Hofman:2006xt} regimes could capture many important features of the worldsheet $S$-matrix of the string theory.}\,
This idea was further developed in \cite{Vazquez:2006hd} by V\'{a}zquez, where it was shown how the giant magnons, or its three-spin generalization, can appear in terms of the string-bit picture.
For the SU(2) sector it was also shown how one can match the canonical structure of the string-bits and the quadratic Hamiltonian of the matrix model with their counterparts in the classical string theory side.

\paragraph{}
In this note, we give further examples of such direct identification between the string theory and the matrix model.
One of the examples of our concern in the string side of the correspondence will be the dyonic giant magnons studied in \cite{Dorey:2006dq,Chen:2006ge,Arutyunov:2006gs,Minahan:2006bd,Spradlin:2006wk,Bobev:2006fg,Kruczenski:2006pk,Okamura:2006zv,Kalousios:2006xy}, and other illustrations will be the limiting cases of the two-spin folded/circular strings of Frolov and Tseytlin \cite{Frolov:2003qc,Frolov:2003xy}.
These solutions will be constructed from dyonic giant magnons, so let us here make some remarks on the dyonic giant magnons.
They are classical string solutions in the so-called Hofman-Maldacena sector \cite{Hofman:2006xt} with two large angular momenta or spins on $S^{5}$.
One of the spins is sent to infinity while the other can be finite, and the energy is also infinite.
More precisely, we can obtain the dyonic giant magnon in the limit
\begin{equation}
E_{\rm DGM}\to \infty\,,\quad 
J_{1}\to \infty\,,\quad 
E_{\rm DGM}-J_{1}\, :\, \mbox{fixed}\,,\quad 
J_{2}\sim \sqrt{\lam}\, :\, \mbox{fixed}\gg 1\,,\quad 
\Delta\varphi \, :\, \mbox{fixed}\,, \quad 
\end{equation}
where $J_{1}$ and $J_{2}$ are the two spins on $S^{5}$, $E_{\rm DGM}$ is the energy, and $\Delta\varphi$ is the angular differene between two endpoints on an equator of the sphere.
The energy-spin relation for the dyonic giant magnon is given by
\begin{equation}
E_{\rm DGM}-J_{1}=\sqrt{J_{2}^{2}+\f{\lambda}{\pi^{2}}\sin^{2}\ko{\f{\Delta\varphi}{2}}}\,.
\label{E-J:DGM}
\end{equation}
We will see in this two-spin generalized case also, as in the single-spin case, we can utilize the string-bit picture to describe the classical string solution.

The crucial idea is to allow the eigenvalues (except two) of background matrices that form string-bits to reside in the interior of the circular droplet.
In other words, while we still take the Hofman-Maldacena limit, we allow the quasi-momentum of each magnon in the SYM state to be complex in general. 
The two eigenvalues on the edge of the droplet represent the ``BPS condensates'' \cite{Vazquez:2006hd} with infinite number of background matrices, and the SYM states dual to the dyonic giant magnons should end in the very two eigenvalues when described as a chain of string-bits.

We will see how this picture works in detail later in Section \ref{sec:Class Strings from MM}.
Before doing so, it would be convenient to review relevant aspects of the earlier works of \cite{Berenstein:2005jq,Vazquez:2006hd} in the following Section \ref{sec:review}.
In Section \ref{sec:Class Strings from MM}, we will also discuss the cases of elliptic folded/circular strings, and also of rational circular strings.
Section \ref{sec:discussion} will be devote to the summary and discussions.

\section{A Review of Berenstein-Correa-V\'{a}zquez Method\label{sec:review}}

We shall first make a brief review on the method for deriving a dispersion relation for composite operators in $\N=4$ SYM at strong coupling a la BCV \cite{Berenstein:2005jq,Vazquez:2006hd}.
We will restrict ourselves to an SU(2) sector throughout this note.

\subsection{The SU(2) Matrix Quantum Mechanics}

Let us start with describing the SU(2) matrix model of our concern.
It is obtained from $\N=4$ SYM on ${\mathbb R} \times S^3$ whose SU(2) scalar part is defined by the action
\begin{equation}
	S_{{\mathbb R} \times S^3}=\f{1}{g_{\rm YM}^2} \int r^3 dt d\Omega_3 \, \tr
	\kko{
	\f{1}{2}(D_\mu \phi_j)^2-\f{\cR}{12}\phi_j^2+\f{1}{4} [\phi_j,\phi_k][\phi_j,\phi_k]
	}\,.
\end{equation}
Here the suffices $j$ and $k$ run from $1$ to $ 4$, and $r$ is the radius of the $S^3$ with the curvature $\cR=6/r^2$.
The real scalar fields $\phi_{j}$ can be expanded by the harmonic functions on $S^3$, of which we are only concerned with the zero-mode.
We can integrate the angular part and obtain the SU(2) matrix model action, setting $r=2/M$\,,
\begin{equation}
S=\frac{2\pi^2}{g^2_{\rm YM}}\left( \frac{2}{M} \right)^3
\int\! dt\, \tr
\kko{\f{1}{2}(D_{t}X_{j})^{2}-\f{1}{2}\ko{\f{M}{2}}^{2} X_{j}^{2}+\f{1}{4} [X_{j},X_{k}][X_{j},X_{k}]}\,.
\end{equation}
In the above, the matrix fields $X_j(t)$ came from the zero-mode of the scalars $\phi_j(t,\vec{x})$.
From standard traceless, Hermitian SU($N$) generators $T^{m}$ $(m=1,\dots,N^{2}-1)$ and time-dependent SO(4)-vectors $X_{j}^{m}(t)$ ($j=1,\dots,4$) in the adjoint representation of SU($N$), we can represent the matrix fields as $(X_{j})^{r}_{s}=X_{j}^{m}(T^{m})^{r}_{s}$\,. 
Particular choice of the parameter $M=2$ (or $r=1$) and redefinition of the fields $\frac{\sqrt{2}\pi}{g_{\rm YM}}\left( \frac{2}{M} \right)^{3/2} X_j \to X_j$ take the action to
\begin{equation}
	S=\int \! dt \, \tr 
	\kko{\f{1}{2}(D_{t}X_{j})^{2}-\f{1}{2}X_{j}^{2}+\f{g_{\rm YM}^2}{8\pi^2}[X_{j},X_{k}][X_{j},X_{k}]}\,.
	\label{eq:action}
\end{equation}
From this action we can obtain the Hamiltonian
\begin{equation}
H=H_{0}+V\,,
\label{eq:Ham}
\end{equation}
where the free part $H_{0}$ and the potential $V$ are given by, respectively,
\begin{equation}
H_{0}=\f{1}{2}\tr (\Pi_{j})^{2}+\f{1}{2}\tr (X_{j})^{2}\,,\qquad 
V=-\frac{g_{\rm YM}^2}{8\pi^2}\tr [X_{j},X_{k}][X_{j},X_{k}]\,.
\end{equation}
Introducing complex scalar fields $Z=\f{1}{\sqrt{2}}(X_1+iX_2)$, $W=\f{1}{\sqrt{2}}(X_3+iX_4)$ and the canonical momenta $\Pi_{Z}=\f{1}{\sqrt{2}}(\Pi_1+i\Pi_2)$, $\Pi_{W}=\f{1}{\sqrt{2}}(\Pi_3+i\Pi_4)$, we can rewrite the Hamiltonian as
\begin{equation}
	H_0=\tr\ko{|\Pi_{Z}|^2+|\Pi_{W}|^2+|Z|^2+|W|^2}\,, \qquad 
	V=\f{g_{\rm YM}^2}{2\pi^2} \tr \ko{|[Z,W]|^2 }\,,
	\label{H_0 and V}
\end{equation}
where we ignored the D-term $\tr\ko{|[Z,\overline{Z}]|^2+|[W,\overline{W}]|^2+|[Z,\overline{W}]|^2}$ in the potential.
\paragraph{}
One can obtain the effective Hamiltonian for $W$ field by treating $Z$ as the normal matrix background, that is, impose the commutation relation $[Z, \overline{Z}]=0$\,.
The Hamiltonian (\ref{eq:Ham}) is invariant under the SU($N$) transformations $Z \to U Z U^\dagger$ and $W \to U W U^\dagger$, and we can use this symmetry to diagonalize the background matrix $Z$ as
\begin{equation}
	Z \, \to\,  U Z U^\dagger =\diag (z_1,\ldots,z_N).
\end{equation}
The same gauge transformation takes $W$ fields to $U W U^\dagger$, which we will again dente as $W$.
The effective Hamiltonian for $W$ is then follows from the transformed potential
\begin{equation}
	V=\f{g_{\rm YM}^2}{2\pi^2} \sum_{i,j=1}^N |z_i-z_j|^2 \,W^i_j \overline{W}_i^j\,,
\end{equation}
as
\begin{equation}
	H_{W}=2\sum_{i,j=1}^N 
	\kko{
	\f{1}{2} (\Pi_{W})^i_j (\overline{\Pi}_{W})^j_i+\f{1}{2} \omega_{ij}^2\, W^i_j \overline{W}_i^j
	}
	=\sum_{i,j=1}^N \omega_{ij} \kko{ (\cW^\dagger)^i_j (\cW)^j_i+(\overline{\cW}^\dagger)^i_j(\overline{\cW})^j_i },
	\label{eq:Ham2}
\end{equation}
where we defined the (effective) creation-annihilation operators for $W$ as
\begin{align}
	(\cW^\dagger)^i_j&=\sqrt{\f{\omega_{ij}}{2}}
	\ko{
	W^i_j-i \f{(\Pi_{W})^i_j}{\omega_{ij}}
	}\,,\quad
	(\cW)^i_j=\sqrt{\f{\omega_{ij}}{2}}
	\ko{
	\overline{W}^i_j+i \f{(\overline{\Pi}_{W})^i_j}{\omega_{ij}}
	}\,,
	\label{cW}\\
	(\overline{\cW}^\dagger)^i_j&=\sqrt{\f{\omega_{ij}}{2}}
	\ko{
	\overline{W}^i_j-i \f{(\overline{\Pi}_{W})^i_j}{\omega_{ij}}
	}\,,\quad
	(\overline{\cW})^i_j=\sqrt{\f{\omega_{ij}}{2}}
	\ko{
	W^i_j+i \f{(\Pi_{W})^i_j}{\omega_{ij}}
	}\,,
	\label{bar cW}
\end{align}
with the frequency
\begin{equation}
	\omega_{ij}=\sqrt{1+\f{g_{\rm YM}^2}{2\pi^2}|z_i-z_j|^2}\,\,.
	\label{omega}
\end{equation}
In deriving (\ref{eq:Ham2}), the zero-point energy was supposed to be canceled by the fermions.
The creation-annihilation operators (\ref{cW}) and (\ref{bar cW}) satisfy the commutation relations
\begin{align}
	&[(\cW)^i_j, (\cW^\dagger)^k_l]=[(\overline{\cW})^i_j, (\overline{\cW}{}^\dagger)^k_l]= \delta^i_l \delta^k_j\,,\quad
	\mbox{otherwise $0$}\,,
	\label{eq:com}
\end{align}
and any SU(2) operators are in one-to-one correspondence with operators that act on the Hilbert space defined by the Hamiltonian (\ref{eq:Ham2}), {\it e.g.},
\begin{equation}
	\tr (Z Z W Z W Z \cdots) \tr (W Z Z Z \cdots) \quad \leftrightarrow\quad 
	\tr (zz \cW^\dagger z \cW^\dagger z \cdots) \tr (\cW^\dagger zzz \cdots) \ket{0}_{W}
\end{equation}
where $\ket{0}_{W}$ is the Fock vacuum for $W$, {\it i.e.}, $\cW \ket{0}_{W}=\overline{\cW} \ket{0}_{W}=0$.

We define the expectation value of the operator ${\mathcal O}$ for the state $\ket{\phi}$ as
\begin{equation}
	\kbra{\phi} {\mathcal O} \kket{\phi}
	\equiv \f{  {\ds \int} \prod\limits_j d^2 z_j |\psi_0(\{z_j \})|^2 \kbra{\phi} {\mathcal O} \kket{\phi}_{W}  }
	{  {\ds \int} \prod\limits_j d^2 z_j |\psi_0(\{ z_j \})|^2 \ip{\phi}{\phi}_{W}  }
	\label{eq:expec}
\end{equation}
where $\psi_0(\{ z_j \})$ is the wavefunction for the fermionic ground state,
\begin{equation}
	\psi_0(\{z_j \})=\prod_{l<k}(z_l-z_k) \exp \Big[-\sum_k |z_k|^2\Big]\,,
	\label{eq:psi0}
\end{equation}
and $\kbra{\phi} {\mathcal O} \kket{\phi}_{W}$ means the expectation value with respect to the operator $W$.
In (\ref{eq:psi0}), the Vandermonde determinant came from the change of integration variables from the original matrix $Z$ to the eigenvalues $\{z_j \}$.
It is well-known the distribution of the eigenvalues $\{ z_j \}$ for the 1/2 BPS ground state in the large $N$ limit is given by a circular droplet, and it can be shown the radius $r_0$ of the droplet is $\sqrt{N/2}$ in our normalization.
Therefore in the large $N$ limit, the integration over $\{ z_{j}\}$ in (\ref{eq:expec}) can be re-expressed as the integration over the droplet,
\begin{equation}
	\kbra{\phi} {\mathcal O} \kket{\phi}
	= \f{  {\ds \int_D} \prod\limits_j d^2 z_j \kbra{\phi} {\mathcal O} \kket{\phi}_{W}  }
	{  {\ds \int_D} \prod\limits_j d^2 z_j \ip{\phi}{\phi}_{W}  }
\end{equation}
where $D$ stands for the circular droplet with radius $r_0=\sqrt{N/2}$.

\subsection{BMN Strings and Giant Magnons as String-Bits}

The BCV method ignores higher order interactions in the potential (\ref{H_0 and V}), which implies the method is only valid for large value of the 't Hooft coupling $\lam$\,.
There are two interesting limit we can consider in this setup: One is the BMN limit \cite{Berenstein:2003gb} where the SYM single-trace composite operator is made up of many $Z$s and few $W$s, see (\ref{BMN limit}) below. The other is the recently invented Hofman-Maldacena limit \cite{Hofman:2006xt}, where the number of $Z$s goes to strictly infinity so that we can relax the trace condition, while $\lam$ is kept fixed, see (\ref{HM limit}).
The classical strings dual to these states are called giant magnons.
As discussed in \cite{Berenstein:2005jq,Vazquez:2006hd}, both the BMN strings and the giant magnons can be expressed in terms of the string-bits in the matrix model.
Below we will briefly review them in turn, which would help us generalize the idea to what we will call ``bound string-bits'' picture later.

\begin{figure}[tb]
\begin{center}
\vspace{.5cm}
\hspace{-.0cm}\includegraphics[scale=0.75]{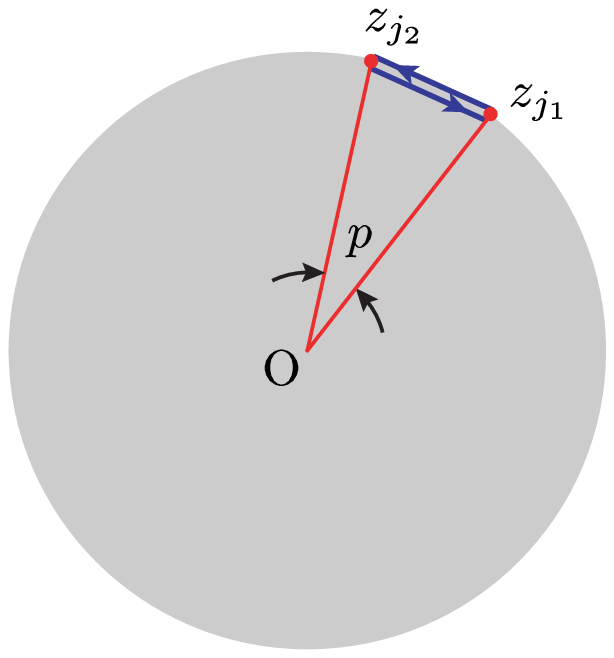}
\hspace{2.0cm}\includegraphics[scale=0.75]{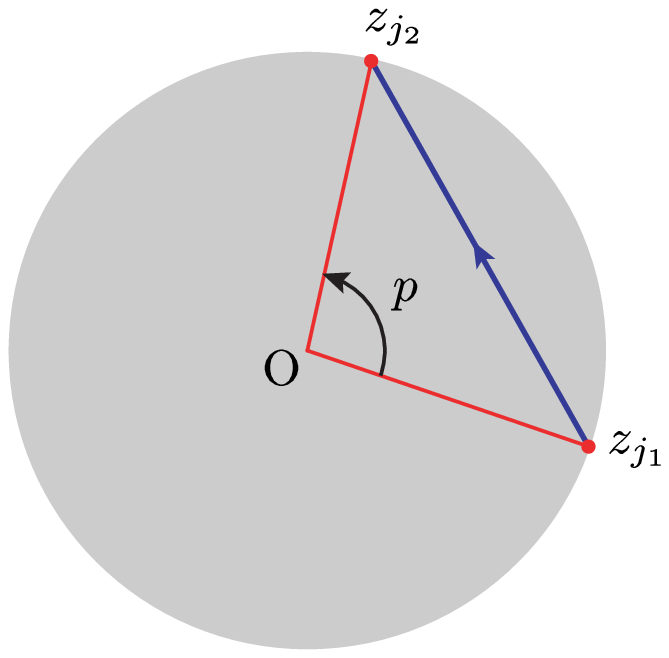}
\vspace{.0cm}
\caption{\small A BMN string (Left) and a giant magnon (Right) as string-bits on a circular droplet of radius $r_{0}=\sqrt{N/2}$.
In both cases the arrow indicates the sign of quasi-momenta, $\exp(ip)=z_{j_{2}}/z_{j_{1}}$\,.}
\label{fig:BMN+GM}
\end{center}
\end{figure}

\paragraph{BMN strings.}
First let us review the BMN case.
The corresponding states in the matrix model is given by
\begin{align}
\ket{p} = \sum_{l=0}^{L} e^{ipl}
\sum_{j_{1},j_{2}}
(z_{j_{1}})^{l} 
(\cW^{\dag})^{j_{1}}_{j_{2}} 
(z_{j_{2}})^{L-l}
(\cW^{\dag})^{j_{2}}_{j_{1}} 
\ket{0}_{W}\,,
\label{|p>:BMN}
\end{align}
where the two magnons or string-bits have real quasi-momenta $\pm p$.
The energy of the BMN states (\ref{|p>:BMN}) is evaluated as
\begin{equation}
	E^{\rm tot.}=L+ \left\langle E^{\rm osc.} \right\rangle\,.
\end{equation}
Here $L$ is the contribution from the background field $Z$ which is supposed to be very large $\sim\sqrt{\lam}$, while the second term is the average energy of the effective Hamiltonian (\ref{eq:Ham2}) with respect to the state (\ref{|p>:BMN}),
\begin{equation}
\left\langle E^{\rm osc.} \right\rangle = \kbra{p}H_{W}\kket{p}
=\f{  {\ds \int_D }\prod\limits_j d^2 z_j \, \kbra{p} H_{W} \kket{p}_{W}  }
	{  {\ds \int_D} \prod\limits_j d^2 z_j \, \ip{p}{p}_{W}  }\,.
\label{eq:E^osc_BMN}
\end{equation}
By using the commutation relation (\ref{eq:com}) for $\cW^\dagger$ and $\cW$, explicitly it reads
\begin{align}
	\left\langle E^{\rm osc.} \right\rangle =
	\f{  {\ds \int_D } \prod\limits_{j} d^2 z_{j} \sum\limits_{j_1,j_2} 2 \omega_{j_1 j_2} 
	\abs{\sum\limits_{l=0}^L\ko{e^{ip} \f{z_{j_1}}{z_{j_2}}}^l z_{j_2}^L }^2
	  }
	{  {\ds \int_D} \prod\limits_{j} d^2 z_{j} \sum\limits_{j_1,j_2} \abs{\sum\limits_{l=0}^L\ko{e^{ip} \f{z_{j_1}}{z_{j_2}}}^l z_{j_2}^L }^2  }\,.
	\label{<E>:BMN}
\end{align}
In the above, we have the factor 2 in the numerator since there are two string-bits joining the eigenvalues $z_{j_{1}}$ and $z_{j_{2}}$.
We evaluate the integrations over the eigenvalues by using a saddle point approximation just as was done in \cite{Berenstein:2005jq}.
Then it is easy to see the sum over $l$ squared in (\ref{<E>:BMN}) is sharply peaked when $z_{j_1}$ and $z_{j_2}$ are related to the magnon quasi-momentum $p$ as
\begin{equation}
e^{ip}=\f{z_{j_2}}{z_{j_1}}\,.
\label{SPC_BMN}
\end{equation}
Moreover, since we are taking $L\to \infty$, both eigenvalues $z_{j_1}$ and $z_{j_2}$ will localize on the edge of the droplet, and become ``BPS condensates'' \cite{Vazquez:2006hd} with infinitely many $Z$s, {\it i.e.}, $\left| z_{j_1} \right|=\left| z_{j_2} \right|=r_{0}$ so that the length of the string-bit becomes $|z_{j_1}-z_{j_2}|=2 r_0 \sin \ko{p/2}$\,. See the Left of Figure \ref{fig:BMN+GM} for the diagram.
Taking all into consideration, we obtain
\begin{equation}
	\left\langle E^{\rm osc.} \right\rangle = 2 \sqrt{1+\f{g_{\rm YM}^2}{2\pi^2} \ko{2r_0 \sin \ko{\f{p}{2}}}^2}
	=2\sqrt{1+\f{\lambda}{\pi^2}\sin^2 \ko{\f{p}{2}}}\,,
\end{equation}
where we have used $r_{0}=\sqrt{N/2}$ and $\lambda= g_{\rm YM}^{2}N$\,.
Finally, taking the BMN limit,
\begin{equation}
L\to \infty\,,\quad 
\lam\to \infty\,,\quad 
p \sim \f{2\pi n}{L} \to 0\,, \quad 
n \, :\, \mbox{fixed}\,, \quad 
\f{\lam}{L^{2}}\, :\, \mbox{fixed}\ll 1\,.
\label{BMN limit}
\end{equation}
we arrive at the famous BMN formula \cite{Berenstein:2003gb} with mode number $n$,
\begin{equation}
E^{\rm tot.}
=L+2\sqrt{1+\f{n^{2}\lam}{L^{2}}}\,.
\end{equation}

\paragraph{Giant magnons.}
Next let us see another interesting example known, that is the giant magnon. 
The relevant state takes the form
\begin{align}
\ket{p} = \sum_{x=-L/2}^{L/2} e^{ipx}
\sum_{j_{1},j_{2}}
(z_{j_{1}})^{(L/2)+x} 
(\cW^{\dag})^{j_{1}}_{j_{2}} 
(z_{j_{2}})^{(L/2)-x}\ket{0}_{W}\,.
\label{|p>:GM}
\end{align}
This corresponds to a SYM state with a single magnon propagating in sea of infinitely many $Z$s, and the asymptoticity allows us to consider a state with nonzero quasi-momentum $p\neq 0$.
See the Right of Figure \ref{fig:BMN+GM} for the diagram.

We can evaluate the energy of the state (\ref{|p>:GM}) in much the same way as the BMN strings.
In this giant magnon case we have for the average energy
\begin{equation}
	\left\langle E^{\rm osc.} \right\rangle =
	\f{  {\ds \int_D } \prod\limits_{j} d^2 z_{j} \sum\limits_{j_1,j_2} \omega_{j_1 j_2} 
	\abs{\sum\limits_{x} z_{j_1}^{L/2} \ko{e^{ip} \f{z_{j_1}}{z_{j_2}}}^x z_{j_2}^{L/2} }^2
	  }
	{  {\ds \int_D} \prod\limits_{j} d^2 z_{j} \sum\limits_{j_1,j_2} \abs{\sum\limits_{x} z_{j_1}^{L/2} \ko{e^{ip} \f{z_{j_1}}{z_{j_2}}}^x z_{j_2}^{L/2} }^2  }\,.
\end{equation}
Again, the saddle point condition (\ref{SPC_BMN}) leads to the energy formula
\begin{align}
	E^{\rm tot.}
	=L+\sqrt{1+\f{\lambda}{\pi^2}\sin^2 \ko{\f{p}{2}}}\,.
	\label{<E>:GM}
\end{align}
If we further take the following Hofman-Maldacena limit,
\begin{equation}
L\to \infty\,,\quad 
\lam\, :\, \mbox{fixed}\gg 1\,,\quad 
p \, :\, \mbox{fixed}\,,\quad 
E^{\rm tot.}-L \, :\, \mbox{fixed}\,,
\label{HM limit}
\end{equation}
then the energy formula (\ref{<E>:GM}) reduces to the dispersion relation for the giant magnon \cite{Hofman:2006xt} having an infinite energy $E_{\rm GM}$ and an infinite angular momentum $J_{1}$ on an equator of $S^{2}$ with angular difference of the two endpoints $\Delta\varphi$,
\begin{equation}
E_{\rm GM}-J_{1}=\f{\sqrt{\lam}}{\pi}\abs{\sin\ko{\f{\Delta\varphi}{2}}}\,.
\label{E-J:GM}
\end{equation}
under the identifications $E^{\rm tot.}\eq E_{\rm GM}$\,, $L\eq J_{1}$ and $p\eq \Delta\varphi$.
\paragraph{}
Thus far we have seen two successful examples of direct correspondence between classical strings and string-bits of the matrix model.
They enabled us to encode, in addition to the spacetime geometry of string theory, the excitations on it into the reduced matrix quantum mechanical model obtained from the dual gauge theory.

Let us summarize the points on what we have seen and what is already known about the giant magnons vs.\,the (elementary) string-bit.
In the SU(2) sector of the correspondence, the edge of the circular droplet in the matrix model setup can be identified with the equatorial circle of the two-sphere on which the string rotates.
The giant magnon can be identified with the single string-bit that joins the two eigenvalues $z_{j_{1}}$ and $z_{j_{2}}$ on the edge, and the string energy is just the Euclidean distance between them \cite{Hofman:2006xt}.
The fact that the endpoints of the giant magnon on the equator carry an infinite spin along the equatorial circle corresponds to the infinitely many background matrices in (\ref{|p>:GM}), and the angular distance between the string endpoints are identified with the magnon momentum.
It would be then natural to ask, when we consider the dyonic giant magnon case, where in the droplet picture the second spin $J_{2}$ enters in.
In the next section, we will give the answer for it.
We will also discuss how folded/circular spinning string solutions of \cite{Frolov:2003qc,Frolov:2003xy} (in a special infinite spin limit) would emerge from our matrix model as collections of many string-bits.

\section{Infinite Spin Limit of Classical Strings from Matrices\label{sec:Class Strings from MM}}

In \cite{Minahan:2006bd}, the authors studied not only the dyonic giant magnons but also a special infinite spin limit of folded and circular strings on ${\mathbb R}\times S^{3}\subset AdS_{5}\times S^{5}$\,.
It was shown that the energy-spin relation for those solutions takes the following universal form
\begin{equation}
E-J_{1}=\sqrt{J_{2}^{2}+k^{2}\lam}\,,\qquad 
E\,,\, J_{1}\to \infty\,,\qquad 
J_{2}\,,\, \lam\,:\,\mbox{fixed}\,,
\end{equation}
with a solution-dependent constant $k$\,.
In the following subsections, we will see how one can reproduce such special solutions via the matrix model metod.

\subsection{Dyonic Giant Magnons as Bound String-Bits\label{sec:DGM from MM}}

So far we have discussed the cases where the SYM states had a single magnon with a real quasi-momentum.
In contrast to them, when we consider states with complex quasi-momenta, it is possible they form a kind of boundstate.
In the familiar Bethe ansatz approach to the SYM spin-chain, boundstates can be defined by the pole condition for the $S$-matrix \cite{Dorey:2006dq}, and the scattering phase-sfhit of the boundstates in the SU(2) sector of the spin-chain was shown to match precisely with the one defined in the classical string theory side under the right gauge choice \cite{Chen:2006gq,Roiban:2006gs}.
What we are going to investigate now is also such magnon boundstates that each constituent magnon has a complex quasi-momentum in general.
However, they are different from the usual boundstates that are related to the poles of $S$-matrices of an integrable spin-chain in that they will be defined without the apparent need of such integrable structure.
See also the discussion in Section \ref{sec:discussion}.
Thus we will refer to the chain of string-bits that minimizes the expectation value of the Hamiltonian (\ref{eq:Ham2}) as a ``bound string-bits'' rather than a boundstate.

\begin{figure}[tb]
\begin{center}
\vspace{.5cm}
\hspace{-.0cm}\includegraphics[scale=0.75]{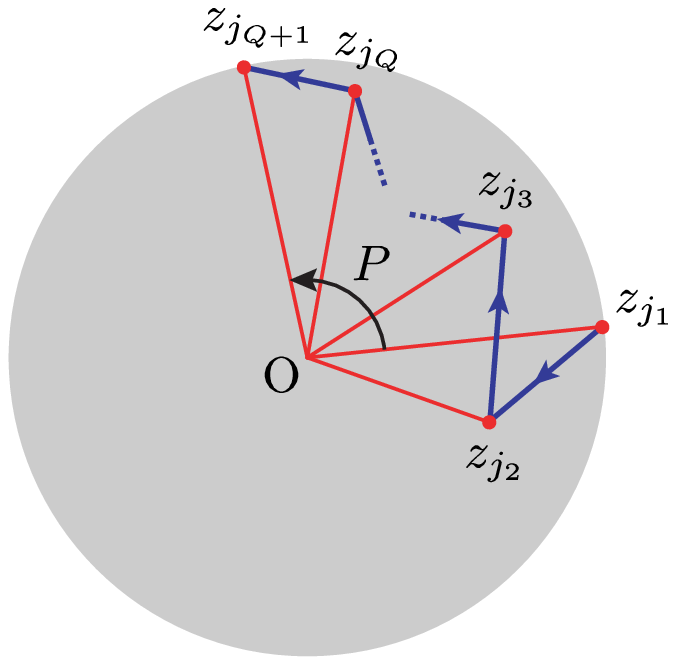}
\hspace{2.0cm}\includegraphics[scale=0.75]{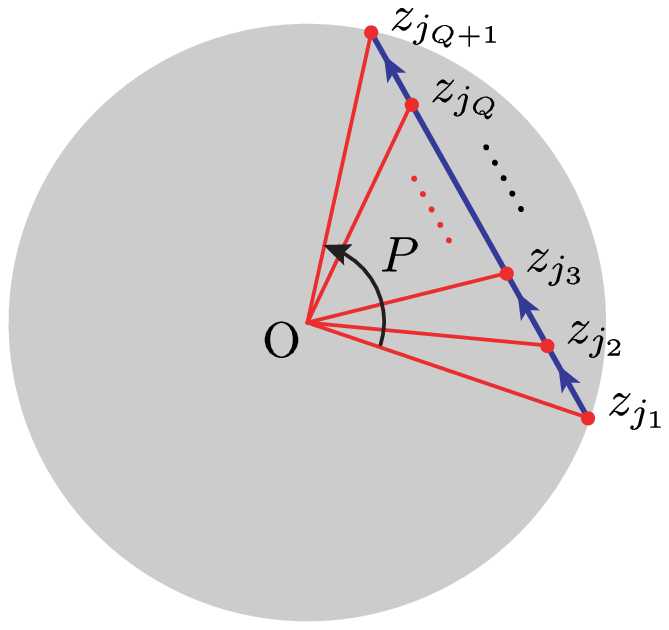}
\vspace{.0cm}
\caption{\small Left diagram shows the generic state while the right shows the lowest energy state for given total quasi-momenta $P$ and the number of string-bits $Q$. 
For the latter case, all eigenvalues $\{ z_{j_{m}} \}$ are equally spaced on one and the same line segment joining two eigenvalues on the edge of the droplet of radius $r_{0}=\sqrt{N/2}$\,.
In the main text we also use abbreviated notations $z_{m}\eq z_{j_{m}}$\,.}
\label{fig:bound_string_bits}
\end{center}
\end{figure}

One might wonder such bound string-bits would require us to take into account the finite $n_{i}$ effect of the BPS condensates $Z^{n_{i}}$ in the SYM states, which would make the problem much harder to tackle on.
To avoid such finite cluster effects, we will again work in the strict Hofman-Maldacena limit where we can ignore the finite $L$ corrections.
Since the ``length'' of the state is infinite in this limit, the number of $Z$s between two adjacent $W$s at sites $x_{i}$ and $x_{i+1}$, that is $n_{i}$, can take values from zero to infinity as we will see below, and it is this feature that still makes in this many-magnon case also some ``classical'' configurations of string-bits possible. 
Our aim here is to show how we can define such bound string-bits in the reduced matrix quantum mechanics setup, especially the state that should be dual to the dyonic giant magnons of string side.
\paragraph{}
Let us start with defining the relevant states.
The $Q$-magnon state in momentum basis will take the following Fourier-transformed form
\begin{align}
\ket{p_{1}, \dots, p_{Q}} &= \sum_{x_{1}<\dots <x_{Q}}\exp\Big(i\sum_{a=1}^{Q}p_{a}x_{a}\Big)
\ket{x_{1}, \dots, x_{Q}}
\label{|p>:DGM}\,,\\
&\hspace{-1.0cm}\mbox{with }\qquad \ket{x_{1}, \dots, x_{Q}}=
\sum_{j_{1},\dots, j_{Q+1}}
(z_{j_{1}})^{n_{1}+x_{1}} 
(\cW^{\dag})^{j_{1}}_{j_{2}} 
(z_{j_{2}})^{x_{2}-x_{1}-1}
\dots \cr
{}&\hspace{3.0cm}
\dots
(z_{j_{Q}})^{x_{Q}-x_{Q-1}-1}
(\cW^{\dag})^{j_{Q}}_{j_{Q+1}} 
(z_{j_{Q+1}})^{n_{Q+1}-x_{Q}}\ket{0}_{W}\,.
\end{align}
The exponents of the two outmost eigenvalues are supposed to scale as $L$, and we explicitly set them as
\begin{equation}
n_{1}=n_{Q+1}=\f{1}{2}\ko{L+Q-1}\,.
\end{equation}
The state (\ref{|p>:DGM}) is simply a generalization of a BMN or a single-spin giant magnon case to a many-magnon case, only now with a set of complex momenta $\{ p_{a} \}$.
As in the giant magnon case studied in the previous section, since we consider asymptotic states ({\it i.e.,} $L \to \infty$), the trace condition is relaxed so that states with non-zero total momentum is allowed.
The average energy of the state (\ref{|p>:DGM}) can be written as $E^{\rm tot.} = L+\left\langle E^{\rm osc.}\right\rangle$ with 
\begin{equation}
\left\langle E^{\rm osc.} \right\rangle = \bra{\{p_{a}\}}H_{W}\ket{\{p_{a}\}}
=\f{  {\ds \int_D }\prod\limits_j d^2 z_j \, \kbra{ \{p_{a}\} } H_{W} \kket{\{p_{a}\}}_{W}  }
	{  {\ds \int_D} \prod\limits_j d^2 z_j \, \ip{\{p_{a}\}}{\{p_{a}\}}_{W}  }\,.
\label{eq:E^osc}
\end{equation}
Explicitly we have in the above
\begin{align}
	\kbra{ \{p_{a}\} } H_{W} \kket{\{p_{a}\}}_{W}
	= \sum_{ \{x_a\},\{x'_a\} }\exp
	\Big[	i\sum_a (-\overline{p}_a x'_a+p_a x_a) \Big] \kbra{ \{x'_{a}\} } H_{W} \kket{\{x_{a}\}}_{W}\,,
	\label{eq:<p|H|p>}
\end{align}
where
\begin{align}
	\kbra{ \{x'_{a}\} } H_{W} \kket{\{x_{a}\}}_{W}
	= &\sum_{ \{j_m\},\{j'_m\} }(\omega_{j_1 j_2}+\omega_{j_2 j_3}+\dots+ \omega_{j_Q j_{Q+1}})\times {} \no \\
	&{}\quad \times (z_{j_1})^{n_1+x_1} (z_{j_2})^{x_2-x_1-1} \dots (z_{j_Q})^{x_{Q}-x_{Q-1}-1} (z_{j_{Q+1}})^{n_{Q+1}-x_Q}\times {} \no \\
	&{}\quad \times (\overline{z}_{j'_1})^{n_1+x'_1} (\overline{z}_{j'_2})^{x'_2-x'_1-1} \dots (\overline{z}_{j'_Q})^{x'_{Q}-x'_{Q-1}-1} (\overline{z}_{j'_{Q+1}})^{n_{Q+1}-x'_Q}\times {} \no \\
	&{}\quad \times \,_W \kbra{0} (\cW)^{j'_2}_{j'_1} (\cW)^{j'_3}_{j'_2} \dots (\cW)^{j'_{Q+1}}_{j'_Q}
	(\cW^\dagger)^{j_1}_{j_2} (\cW^\dagger)^{j_2}_{j_3} \dots (\cW^\dagger)^{j_{Q}}_{j_{Q+1}} \kket{0}_{W} \no \\
	=& \sum_{ \{j_m\} } (\omega_{j_1 j_2}+\omega_{j_2 j_3}+\dots+ \omega_{j_Q j_{Q+1}}) \times {}\no \\
	&{}\quad \times (z_{j_1})^{n_1+x_1} (z_{j_2})^{x_2-x_1-1} \dots (z_{j_Q})^{x_{Q}-x_{Q-1}-1} (z_{j_{Q+1}})^{n_{Q+1}-x_Q} \times {}\no \\
	&{}\quad \times (\overline{z}_{j_1})^{n_1+x'_1} (\overline{z}_{j_2})^{x'_2-x'_1-1} \dots (\overline{z}_{j_Q})^{x'_{Q}-x'_{Q-1}-1} (\overline{z}_{j_{Q+1}})^{n_{Q+1}-x'_Q}
	+\dots\,.
	\label{eq:<x|H|x>}
\end{align}
In (\ref{eq:<x|H|x>}), the last dots represents terms that vanish when we integrate over $z$, and in what follows we will omit them. 
Substituting (\ref{eq:<p|H|p>}) and (\ref{eq:<x|H|x>}) into (\ref{eq:E^osc}), we obtain
\begin{equation}
	\left\langle E^{\rm osc.} \right\rangle =\sum_{m=1}^{Q}\ep_{m}
	\label{eq:E^osc2}
\end{equation}
where $\epsilon_m$ is the contribution from the string-bit joining $m$-th and $(m+1)$-th eigenvalues,
\begin{equation}
\ep_{m}=\f{{\ds \int_D}\prod\limits_{j}d^{2}z_{j} \sum\limits_{ \{j_n\} } \omega_{j_{m} j_{m+1}}\left| \sum\limits_{\{x_{a}\}} (z_{j_{1}})^{n_{1}}\prod\limits_{k=1}^{Q}\kko{\ko{e^{ip_{k}}\f{z_{j_{k}}}{z_{j_{k+1}}}}^{x_{k}}}(z_{j_{Q+1}})^{n_{Q+1}}
 \right|^{2}}{{\ds \int_D}\prod\limits_{j}d^{2}z_{j} \sum\limits_{ \{j_n\} } \left| \sum\limits_{\{x_{a}\}} (z_{j_{1}})^{n_{1}}\prod\limits_{k=1}^{Q}\kko{\ko{e^{ip_{k}}\f{z_{j_{k}}}{z_{j_{k+1}}}}^{x_{k}}}(z_{j_{Q+1}})^{n_{Q+1}}
 \right|^{2}}\,.
 \label{ep_m}
\end{equation}
We evaluate these integrals over $z$ in much the same way as in the previous section.
For notational simplicity, let us introduce abbreviated notations 
$p_{m}\eq p_{j_{m}}$ and 
$z_{m}\eq z_{j_{m}}$.
The saddle point (stationary phase) conditions for this many-magnon case become
\begin{equation}
e^{ip_{k}}\f{z_{k}}{z_{k+1}}=1 \quad {\it i.e.}, \quad
e^{ip_{k}}=\f{z_{k+1}}{z_{k}}
\qquad \mbox{for}\quad 
k=1,\dots,Q\,.
\label{eq:SP}
\end{equation}
Since we are working in the limit $L\to \infty$, the outmost eigenvalues $z_1$ and $z_{Q+1}$ are localized on the edge of the circular droplet, which implies the total quasi-momentum $P\eq \sum_{m=1}^{Q}p_{m}$ is real, $|e^{iP}|=|z_{Q+1}|/|z_{1}|=1$\,.
In other words, $P$ is the azimutal angle between $z_1$ and $z_{Q+1}$ in the droplet, see the Left diagram of Figure \ref{fig:bound_string_bits}.
The rest $Q-1$ eigenvalues can reside in the interior of the droplet, $\left| z_{k} \right|\leq r_0$ $(k=2,\dots,Q)$, which reflects the fact the quasi-momenta generally have imaginary parts in our bound string-bit case.

In general, the chain of line segments, or string-bits, joining $z_{m}$ and $z_{m+1}$ $(m=1,\dots,Q)$ successively form an open zig-zag line as shown in the Left of Figure \ref{fig:bound_string_bits}. 
Let us now recall that dyonic giant magnons correspond to BPS boundstates in an asymptotic SYM spin-chain with centrally-extended supersymmetry algebra $({\rm PSU(2|2)\times PSU(2|2)})\ltimes {\mathbb R^{3}}$.
With this in mind, it is plausible in our matrix model case also the corresponding string-bits configuration is associated with a special set of quasi-momenta $\{ p_{a} \}$ which minimizes the total energy of the string-bits.
Such a set of quasi-momenta will determine the locations of eigenvalues $\{ z_{m} \}$ in the droplet uniquely for given number of string-bits (or magnons) $Q$ and the total momentum $P$.
We will soon see this is indeed the case, and the configuration with the lowest energy configuration precisely reproduces the dispersion relation for the dyonic giant magnons.
It can be easily verified that the approximated energy $\left\langle E^{\rm osc.} \right\rangle$ in (\ref{eq:E^osc2}) takes its minimum when
\begin{equation}
\left| z_{1}-z_{2} \right| = \dots = \left| z_{Q}-z_{Q+1} \right| = \f{1}{Q}\left| z_{1}-z_{Q+1}\right| = \f{2r_{0}}{Q}\sin\ko{\f{P}{2}}\,,
\label{z-z}
\end{equation}
that is, all eigenvalues $\{ z_{m} \}$ lie on one and the same line segment joining two points on the circle, and they are all equally spaced (see the Right of Figure \ref{fig:bound_string_bits}).
This condition along with the limit
\begin{equation}
L\to \infty\,,\quad 
Q\sim \sqrt{\lam}\, :\, \mbox{fixed}\gg 1\,,\quad 
P \, :\, \mbox{fixed}\,, \quad 
\end{equation}
leads to the following energy formula
\begin{equation}
\left. E^{\rm tot.} - L\vphantom{\f{}{}}\right|_{\rm min}
=\left\langle E^{\rm osc.} \right\rangle_{\rm min}
=Q\times \sqrt{1+\f{g_{\rm YM}^{2}}{2\pi^{2}}\ko{\f{2r_{0}}{Q}\sin\ko{\f{P}{2}}}^{2}}
=\sqrt{Q^{2}+\f{\lambda}{\pi^{2}}\sin^{2}\ko{\f{P}{2}}}\,,
\label{E-L}
\end{equation}
which precisely reproduces the energy-spin relation for the dyonic giant magnons (\ref{E-J:DGM}) under the identifications $Q\eq J_{2}$\,, $E^{\rm tot.}\eq E_{\rm DGM}$\,, $L\eq J_{1}$, and $P\eq \Delta\varphi$.
It also matches with the exact BPS dispersion relation for $Q$-magnon boundstates in the asymptotic $\N=4$ SYM spin-chain, which can be obtained from purely group theoretical means \cite{Chen:2006gp}.\footnote{\,Supersymmetry alone can only determine the form of the dispersion to be $\Delta-J_{1}=\sqrt{J_{2}^{2}+f(\lam)\sin^{2}\ko{\f{P}{2}}}$, and one has to take into account the known perturbative results of SYM theory to obtain $f(\lam)=\lam/\pi^{2}$ (which is known to be vald at least up to the three-loop order), which matches with (\ref{E-J:DGM}) or (\ref{E-L}).}\,
Note also that we could have considered $Q\ll \sqrt{\lam}$ region with $\lam$ large but finite.
In this case the dispersion relation (\ref{E-L}) reduces to the one for the single-spin giant magnon case (\ref{E-J:GM}), which simply measures the distance between the two endpoint of the straight stick.
\paragraph{}
In summary, we argue that the degree of freedom of the second spin $J_{2}$ of the dyonic giant magnons, which is orthogonal to the first, infinite spin $J_{1}$ along the equator, appears in the matrix model setup as the number of line segments with equal length. 
It is obtained by dividing the straight line joining two eigenvalues on the edge of the circular droplet into $J_{2}$ pieces with equal length as (\ref{z-z}).
The angular difference between the two outmost eigenvalues $P$ corresponds to the angular difference between the endpoints of the dyonic giant magnon $\Delta \vp$, as is the case with the single-spin giant magnon.

\subsection{Infinite Spin Limit of Folded/Circular Strings\label{sec:F/C from MM}}

Having worked out the generalities, we can now straightforwardly apply the arguments to other interesting string solutions, that are the special infinite spin limit of folded and circular spinning strings \cite{Minahan:2006bd}.
Let us parametrize the metric of ${\mathbb R}\times S^{3}$ in $AdS_{5}\times S^{5}$ as
\begin{equation}
ds^{2}_{\,{\mathbb R}\times S^{3}}
=-dt^{2}+d\theta^{2}+\sin^{2}\theta d\vp_{1}^{2}+\cos^{2}\theta d\vp_{2}^{2}\,.
\end{equation}
The folded and circular strings are obtained by imposing appropriate ansatze on $(t,\theta,\vp_{1},\vp_{2})$\,.

\paragraph{\bmt{J_{1}\gg J_{2}} limit of elliptic folded/circular strings.}

First let us see the cases of two-spin elliptic strings.
Among them, two well-known examples are the folded and the circular strings studied in \cite{Frolov:2003qc,Frolov:2003xy}.

The profile of the folded string solution is given by, setting the number of folds one,
\begin{align}
t=\kappa\tau\,,\quad 
\cos\theta(\sigma)={\rm dn}(A\sigma,q)\,,\quad 
\sin\theta(\sigma)=\sqrt{q}\,{\rm sn}(A\sigma,q)\,,\quad 
\vp_{j}=w_{j}\tau\quad (j=1,2)\,,
\end{align}
with $A\eq \sqrt{w_{1}^{2}-w_{2}^{2}}$\,.
Here the elliptic moduli $q$ is related to other parameters as
\begin{equation}
q\eq \sin^{2}\theta_{\ast}=\f{\kappa^{2}-w_{2}^{2}}{w_{1}^{2}-w_{2}^{2}}\,.
\end{equation}
where $\theta_{\ast}$ is a parameter that fixes the range of the folded string as $-\theta_{\ast}\leq \theta\leq \theta_{\ast}$\,.
The energy $E$ and the two spins $J_{1}$ and $J_{2}$ are defined as
\begin{equation}
E=\sqrt{\lam}\,\kappa\,,\quad 
J_{1}=\sqrt{\lam}\,w_{1}\int_{0}^{2\pi}\f{d\sigma}{2\pi}\sin^{2}\theta\,,\quad 
J_{2}=\sqrt{\lam}\,w_{2}\int_{0}^{2\pi}\f{d\sigma}{2\pi}\cos^{2}\theta\,,
\label{global charges}
\end{equation}
and satisfy the following pair of transcendental equations
\begin{align}
\ko{\f{E}{\eK(q)}}^{2}-\ko{\f{J_{2}}{\eE(q)}}^{2}
&=\f{4\lam}{\pi^{2}}\,q\,,\cr
\ko{\f{J_{1}}{\eK(q)-\eE(q)}}^{2}-\ko{\f{J_{2}}{\eE(q)}}^{2}
&=\f{4\lam}{\pi^{2}}\,,
\label{fold}
\end{align}
where $\eK(q)$ and $\eE(q)$ are the standard complete elliptic integrals of the first and the second kind.\footnote{\,Our convention for the complete elliptic integrals of the first and the second kind are as follows:
\begin{alignat}{3}
\eK(q) &\equiv \int_{0}^{1} \frac{d x}{\sqrt{\ko{1-x^2}\ko{1-qx^2}}}&{}&=\int_{0}^{\pi/2} \frac{d \varphi}{\sqrt{1-q\sin^{2}\varphi}}\,,\cr
\eE(q) &\equiv \int_{0}^{1} d x\,\sqrt{\frac{1-qx^2}{1-x^2}}&{}&=\int_{0}^{\pi/2} d \varphi\,\sqrt{1-q\sin^{2}\varphi}\,.\no
\end{alignat}
Note the parameter $A$ in the profile of folded/circular strings can be written as $A=\f{2}{\pi}\eK(q)$\,.
}\,
By eliminating $\eK(q)$ in (\ref{fold}) and taking $q\to 1$ ({\it i.e.}, $\theta_{\ast}\to \pi/2$) limit, where $\eK(q)\to \infty$ and $\eE(q)\to 1$, we arrived at \cite{Dorey:2006dq,Minahan:2006bd}
\begin{equation}
E-J_{1}=\sqrt{J_{2}^{2}+\f{4\lam}{\pi^{2}}}\,.
\label{elliptic}
\end{equation}

It can be shown that the same expression for the energy-spin relation results from the infinite spin limit of the circular solution case, whose profile is given by
\begin{align}
t=\kappa\tau\,,\quad 
\cos\theta(\sigma)={\rm cn}(A\sigma,\tilde q)\,,\quad 
\sin\theta(\sigma)={\rm sn}(A\sigma,\tilde q)\,,\quad 
\vp_{j}=w_{j}\tau\quad (j=1,2)\,.
\end{align}
Here the moduli for the circular case is related to the one in the folded case as $\tilde q=1/q$\,.
Evaluating the circular counterparts of (\ref{fold}) (with the global charges defined as (\ref{global charges}))
\begin{align}
\ko{\f{E}{\eK(\tilde q)}}^{2}-\ko{\f{{\tilde q}J_{2}}{(1-{\tilde q})\eK(\tilde q)-\eE(\tilde q)}}^{2}
&=\f{4\lam}{\pi^{2}}\,,\cr
\ko{\f{{\tilde q}J_{1}}{\eK(\tilde q)-\eE(\tilde q)}}^{2}-\ko{\f{{\tilde q}J_{2}}{(1-{\tilde q})\eK(\tilde q)-\eE(\tilde q)}}^{2}
&=\f{4\lam}{\pi^{2}}\,\tilde q\,,
\label{circ}
\end{align}
one is again end up with the same energy-spin relation as (\ref{elliptic}) in the $\tilde q\to 1$ limit.
This reflects the fact that the limiting behaviors of the folded and the circular strings are identical, or more precisely, the configurations can be switched from one to another without energy cost.
%

\begin{figure}[tb]
\begin{center}
\vspace{.5cm}
\hspace{-.0cm}\includegraphics[scale=0.75]{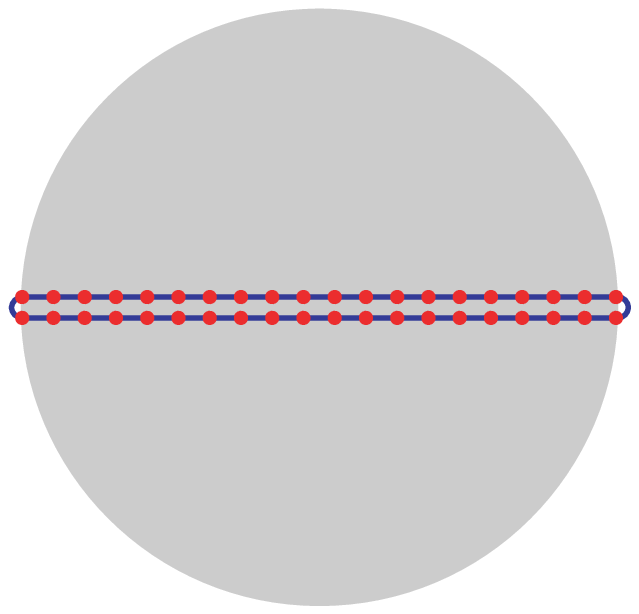}
\hspace{2.0cm}\includegraphics[scale=0.75]{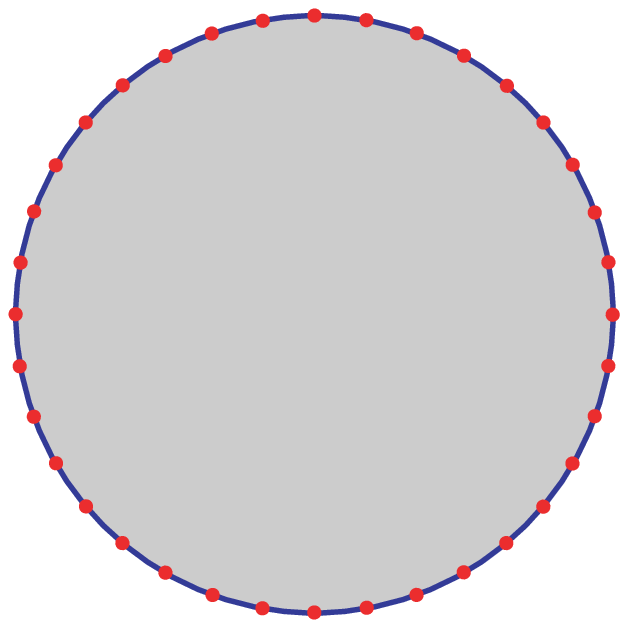}
\vspace{.0cm}
\caption{\small $J_{1}\gg J_{2} \, (\sim \sqrt\lam\gg 1)$ limit of elliptic folded/circular strings (Left) and a rational circular string (Right) as closed chains of string-bits on a circular droplet of radius $r_{0}=\sqrt{N/2}$\,.
The rotation about the origin of the droplet is generated by $J_{1}$ charge, and the number of eigenvalues that form the hain of string-bits is $J_{2}$, which is a continuous variable in classical string theory.}
\label{fig:elliptic+rational}
\end{center}
\end{figure}

Since we have already seen the generic case of dyonic giant magnons in the previous subsection, it is almost trivial to understand how one can reproduce the infinite spin limit of these elliptic folded/circular solutions as a chain of string-bits.
They are made up of two copies of dyonic giant magnons with $P=\pi$, and the corresponding configurations of $J_{2}$ string-bits are given by the Left diagram of Figure \ref{fig:elliptic+rational}.
All the $J_{2}$ eigenvalues are located on one and the same diameter of the circular droplet with equal spacing, and reflecting the closedness of the strings, the string-bits associated with them also form closed chains.
The energy of the closed chain can be evaluated as, noticing that the distance between two adjacent eigenvalues are given by $2r_{0}/(J_{2}/2)$\,,
\begin{equation}
J_{2}\times \sqrt{1+\f{g_{\rm YM}^{2}}{2\pi^{2}}\ko{\f{2r_{0}}{J_{2}/2}}^{2}}
=\sqrt{J_{2}^{2}+\f{4\lam}{\pi^{2}}}\,,
\label{elliptic2}
\end{equation}
which matches with the string theory result (\ref{elliptic}).
For a generic $n$-fold case we just need to multiply $\lam$ in the dispersion relation (\ref{elliptic}) (or (\ref{elliptic2})) by $n^{2}$.

\paragraph{\bmt{J_{1}\gg J_{2}} limit of rational circular strings.}

Next let us see the case of a rational circular (or a simplest circular) string.
This solution follows from the ansatz
\begin{equation}
t=\kappa\tau\,,\quad 
\theta=\theta_{0}=\mbox{const.},\quad 
\vp_{j}=w_{j}\tau + m_{j}\sigma\quad (j=1,2)\,.
\end{equation}
and the global charges are given by
\begin{equation}
E=\sqrt{\lam}\,\kappa\,,\quad 
J_{1}=\sqrt{\lam}\,w_{1}\sin^{2}\theta_{0}\,,\quad 
J_{2}=\sqrt{\lam}\,w_{2}\cos^{2}\theta_{0}\,.
\end{equation}
It can be shown that in $m_{2}/m_{1}\to -\infty$ ($\theta_{0}\to \pi/2$) limit, the energy-spin relation for the limiting rational string becomes \cite{Minahan:2006bd}
\begin{equation}
E-J_{1}=\sqrt{J_{2}^{2}+n^{2}\lam}\,,
\label{rational}
\end{equation}
where we have denoted $n\eq m_{1}$\,.

In this rational circular case also one can obtain a string-bits configuration whose dispersion relation exactly reproduces the energy-spin relation for the corresponding string.
See the Right of Figure \ref{fig:elliptic+rational}.
All the $J_{2}$ eigenvalues that form a closed chain of string-bits are equally spaced on the perimeter of the circular droplet.
Let $n$ be the number of windings along the circuit (which should be identified with the winding number for the circular string), then the energy of the closed chain can be evaluated as
\begin{align}
J_{2}\times \sqrt{1+\f{g_{\rm YM}^{2}}{2\pi^{2}}\ko{2r_{0}\sin\ko{\f{\pi n}{J_{2}}}}^{2}}
=\sqrt{J_{2}^{2}+\f{J_{2}^{2}\lam}{\pi^{2}}\sin^{2}\ko{\f{\pi n}{J_{2}}}}
\, \approx \, 
\sqrt{J_{2}^{2}+n^{2}\lam}\,,
\label{rational from MM}
\end{align}
which precisely reproduces (\ref{rational}) in the limit $J_{2}\gg 1$ with fixed $n$\,.

Note also the $J_{1}\gg J_{2}$ limit of the rational circular string can be regarded as an infinite array of infinitesimal dyonic giant magnons, which is also a limiting case of a so-called ``helical'' string \cite{Okamura:2006zv}.
In fact, in this way we can interpret the configuration of Figure \ref{fig:elliptic+rational} more easily as the classical string.
Let $n$ be the number of windings along the equator of $S^{3}$ when $\sigma$ goes from $0$ to $2\pi$, and $p$ be the angular difference associated with each dyonic giant magnon. 
Then the number of magnons is given by $m=2\pi n/p$\,, and the energy-spin relation for the array of magnons is computed as, denoting its total energy and two spins as $E\, (\to \infty)$, $J_{1}\, (\to \infty)$, and $J_{2}\, (\sim \sqrt{\lam}\mbox{\,:\,fixed})$, respectively, 
\begin{equation}
E-J_{1}=\mathop {\lim \vphantom{{}_{}}}\limits_{\hfill \scriptstyle {\vphantom{\f{}{}}m\to \infty }\hfill\atop
  \scriptstyle {\hfill n\,:\, {\rm fixed}}\hfill}
  \kko{m\times \sqrt{\ko{\f{J_{2}}{m}}^{2}+\f{\lam}{\pi^{2}}\sin^{2}\ko{\f{p}{2}}}\, }
=\sqrt{J_{2}^{2}+n^{2}\lam}\,,
\end{equation}
which matches with (\ref{rational}) and also with (\ref{rational from MM}).

\section{Summary and Discussions\label{sec:discussion}}

In this note, by using the reduced matrix quantum mechanics of \cite{Berenstein:2005jq,Vazquez:2006hd}, we showed the particular distribution of eigenvalues (\ref{z-z}), which we called ``bound string-bits'', can be identified with a classical string known as a dyonic giant magnon.
We proposed the lowest energy configuration of string-bits (\ref{z-z}) under fixed $P$ and $Q$ be the matrix model dual of the dyonic giant magnons, reproducing the dispersion relation as (\ref{E-L}).
We also showed special infinite spin limits of folded/circular strings of elliptic types and also the rational circular string can be described as closed chains of string-bits.

In any case, what was surprising is that the configurations of string-bits can be viewed as the ``shadows'' of the corresponding strings on $S^{5}$ projected onto the equatorial circle, which can be identified with the circular droplet of eigenvalues.\footnote{\,To see this for the dyonic giant magnon case, let us choose a rotating frame on the sphere where the point with infinite $J_{1}$ is stationary, that is, parametrize the profile of the string by ${\widetilde Z}_{1}=e^{-it}Z_{1}$ and $Z_{2}$, where $Z_{j}$ $(j=1,2)$ are the ones used in \cite{Chen:2006ge}, see Eqn (39) in their paper.
Setting ${\widetilde Z}_{1}={\widetilde X}_{1}+i{\widetilde X}_{2}$, we see that $-{\widetilde X}_{2}={k \mathord{\left/ {\vphantom {\sqrt{1+k^{2}}}} \right. \kern-\nulldelimiterspace} \sqrt{1+k^{2}}}=\cos(p/2)$ is constant in $\sigma$ and $\tau$, which means the ``shadow'' of the dyonic giant magnon projected onto the ${\widetilde X}_{1}$-${\widetilde X}_{2}$ plane is just a straight stick connecting two points on the equatorial circle of $S^{3}$, as shown in the Right of Figure \ref{fig:bound_string_bits}.}\,
The center of the circular droplet (or the origin of the LLM coordinate) corresponds to the northern and southern poles of the sphere.
Thus via the BCV approach we could see not only the emergence of geometries \cite{Berenstein:2005jq} but also of further examples of emergent classical strings than ever known.
The BMN strings and giant magnons were already obtained in \cite{Berenstein:2005jq,Vazquez:2006hd}, and in this note we have added to the dictionary two-spin examples: the dyonic giant magnons, the elliptic folded/circular strings and the rational circular strings in the infinite spin limit.
The latter examples can be obtained from the dyonic giant magnons as we have seen in Section \ref{sec:F/C from MM}.

\paragraph{}
Among many possible further directions, it is important to extend the analysis to finite $L$ case, taking into consideration the finite size effect.
Obviously there are two sources for this effect.
One is the correction coming from the approximation we used. 
Recall that in evaluating the energy for each string-bit (\ref{ep_m}), we employed saddle point approximations.
The sum in $|\dots|^{2}$ in (\ref{ep_m}) in the large $L$ limit becomes, roughly, 
\begin{align}
&\left| \sum\limits_{\{x_{a}\}} (z_{j_{1}})^{n_{1}}\prod\limits_{k=1}^{Q}\kko{\ko{e^{ip_{k}}\f{z_{j_{k}}}{z_{j_{k+1}}}}^{x_{k}}}(z_{j_{Q+1}})^{n_{Q+1}}
 \right|^{2}\cr
 &\qquad \qquad {}\sim \quad 
r_{0}^{2(L+Q-1)}\prod\limits_{k=1}^{Q}\abs{\f{\sin\ko{w_{k}L/2}}{\sin\ko{w_{k}/2}}}^{2}\,,
\qquad 
\mbox{where}\quad e^{iw_{k}}\eq e^{ip_{k}}\f{z_{j_{k}}}{z_{j_{k+1}}}\,,
\end{align}
which is the Laue function for $Q$-dimensional crystal lattice.
The approximation that lead to (\ref{eq:SP}) amounts to take account of only the leading Laue peaks in both the denominator and the numerator, and throw away the other smaller peaks.
Taking into account of those smaller peaks will enable us to evaluate the finite size correction to the ``classical'' energy (\ref{E-L}).
The other source is the backreaction of the chain of string-bits to the geometry made up of the background field $Z$.
Since the effect of the backreaction is entangled with the $1/\sqrt{\lambda}$ correction, the quadratic Hamiltonian we started with will fail to capture the correct physics in the finite $L$ region; rather we will have to include in the Hamiltonian the higher order interactions.
It would not be an easy task to take into account all those finite $L$ corrections consistently, but if successfully done, then one might be even able to compare the result with the finite-size corrected (dyonic) giant magnons of \cite{Arutyunov:2006gs,Okamura:2006zv}.
\paragraph{}
It is also interesting to investigate other space-time geometries such as $AdS_{5}\times Y^{p,q}$ in this direction.
Not only the geometries themselves but also the excitations on them are important objects to investigate.
As a simple example, below we will show how giant magnons in Lunin-Maldacena background \cite{Lunin:2005jy} emerges as string-bits in the matrix model.
This background is conjectured to be dual to the special case of $\beta$-deformed $\N=4$ SYM theory of Leigh and Strassler \cite{Leigh:1995ep}, which has $\N=1$ supersymmetry.
For simplicity, take $\beta$ parameter to be real for now, then the potential term in the $\be$-deformed SU(2) sector is simply given by
\begin{equation}
V_{\be}=\f{g_{\rm YM}^2}{2\pi^2} \tr \ko{|[Z,W]_{\be}|^2 }\,,\qquad 
[Z,W]_{\be}\eq ZW-e^{-2\pi i \be}WZ\,.
\end{equation}
In $\be\to 0$ limit, this reduces to the undeformed potential of (\ref{H_0 and V}).
The frequency for each string-bit becomes 
\begin{equation}
	\omega_{j_{m}j_{m+1},\be}=\sqrt{1+\f{g_{\rm YM}^2}{2\pi^2}\abs{z_{j_{m}}-e^{-2\pi i \be}z_{j_{m+1}}}^2}\,,
\end{equation}
and the saddle point condition is given by $e^{i{\tilde p}_{m}}=e^{i(p_{m}+2\pi\be)}=z_{j_{m+1}}/z_{j_{m}}$ for this $\be$-deformed case.
Here $\tilde p_{m}$ denotes the quasi-momentum in the deformed theory and $p_{m}$ does the original one.
We obtain the minimized energy of the chain of $Q$ string-bits,
\begin{equation}
\left. E^{\rm tot.}_{\be} - L\vphantom{\f{}{}}\right|_{\rm min}
=\sqrt{Q^{2}+\f{\lambda}{\pi^{2}}\sin^{2}\ko{\f{P}{2}+\pi Q\be}}\,,
\label{E-L:gamma}
\end{equation}
where $P=\sum_{m=1}^{Q}p_{m}$ as before.
As can be seen in the above, the twisting effect of $\be$ parameter in the potential only results in the shift of the quasi-momentum of the magnon.\footnote{\,The $Q=1$ case was already displayed in \cite{Berenstein:2005ek}.}
This observation agrees with the one made in the Bethe ansatz approach \cite{Frolov:2005ty}, where the corresponding spin-chain was an ${\rm XXX}_{1/2}$ spin-chain with a twisted boundary condition.
The dispersion relation (\ref{E-L:gamma}) can be also compared to the energy-spin relations for the two-spin giant magnons obtained in \cite{Chu:2006ae,Bobev:2006fg}.\footnote{\,By relating the magnon quasi-momentum, the deformation parameter $\be$ and the parameter that determines the string configuration in a particular way, the authors of \cite{Chu:2006ae,Bobev:2006fg} derived a dispersion relation of the form $E-J_{1}=\sqrt{J_{2}^{2}+\f{\lam}{\pi^{2}}\sin^{2}\ko{\f{P}{2}-\pi\be}}$\,. It is different from what we have obtained in the above, except the elementary magnon case $J_{2}=Q=1$.}\,
It is interesting to note that the shifting operation in the TsT-transformation \cite{Frolov:2005dj}, which is an $U(1)$ rotation along one of the isometries of $S^{5}$, can thus be directly identified in the droplet plane or the LLM plane.

\paragraph{}
In closing, we would like to make some comments on integrability issue.
It has been recently shown that dyonic giant magnon corresponds to a magnon boundstate in the Beisert-Dippel-Staudacher (BDS) spin-chain \cite{Beisert:2004hm} of $\N=4$ SYM theory.
Turning back to the matrix models, as is also well-known, the SU(2) matrix model has very similar integrable structure as the mother $\N=4$ theory.
In fact the phase-function for the model is conjectured to be exactly the same as the $\N=4$ SYM $S$-matirx of BDS and also the string $S$-matrix of Arutyunov-Frolov-Staudacher (AFS) \cite{Arutyunov:2004vx}, the difference of those three $S$-matrices (matrix model, BDS, AFS) being only the overall phase factors called dressing factors \cite{Fischbacher:2004iu}.
Therefore it would be natural to expect a possible interpretation of our bound string-bits in terms of magnon dynamics in the $\N=4$ theory.

As in the BDS case, generic $Q$-magnon boundstate in the SU(2) matrix model can be defined by the pole condition for the conjectured $S$-matrix in the matrix model.
It reads $\vp(p_{j})-\vp(p_{j+1})=i$ $(j=1,\dots,Q-1)$ in terms of the phase-function $\vp(p_{j})=\f{1}{2}\cot\ko{\f{p_{j}}{2}}\sqrt{1+\f{\lambda}{\pi^{2}}\sin^{2}\ko{\f{p_{j}}{2}}}$ with the complex quasi-momenta $\{ p_{j} \}$.
Comparing this with the lowest energy distribution of eigenvalues (\ref{z-z}), one might be tempted to expect a close relation between the phase-function variables $\vp(p_{j})$ that form a Bethe string and the eigenvalues $z_{j}$ that form a chain of string-bits.
In fact the condition (\ref{z-z}) coincides with the pole conditions for the $S$-matrix of one-loop gauge theory, which has the same integrability as Heisenberg spin-chain.
At the one-loop level, the BDS phase-function $\vp(p)$ reduces to $\vp_{0}(p)=\f{1}{2}\cot\ko{\f{p}{2}}$, with which the pole condition $\vp_{0}(p_{j})-\vp_{0}(p_{j+1})=i$ $(j=1,\dots,Q-1)$ exactly matches with the condition $z_{1}-z_{2}=\dots = z_{Q}-z_{Q+1}$ up to the degree of freedom of rotating the droplet plane. 
Notice that, however, our ``bound string-bits'' has been defined such that it minimizes the energy of the chain of string-bits under the condition of fixed total quasi-momentum and fixed number of constituent magnons, and it is not, at least apparently, a consequence of any integrablity of the SU(2) matrix model.
In this regard, it is not surprising if it turned out the boundstates defined as poles of $S$-matrix in the SU(2) matrix model or the BDS spin-chain do not correspond to our bound string-bits in a direct manner.
Actually it can be verified that the set of quasi-momenta $\{ p_{j} \}$ that defined the straight stretched line segment joining $z_{1}$ and $z_{Q+1}$ does not satisfy the BDS pole conditions in the strong coupling.

As yet we have no clear answer as to whether our bound string-bits can be defined by some requirement related to integrability.
We hope to report on this as another publication in the future.

\subsubsection*{Acknowledgments}

The authors would like to thank H.-Y.~Chen, D.~H.~Correa, R.~Suzuki and S.~E.~V\'{a}zquez for their valuable comments and discussions.



\begin{thebibliography}{99}

\bibitem{Maldacena:1997re}
  J.~M.~Maldacena,
  ``The large $N$ limit of superconformal field theories and supergravity,''
  Adv.\ Theor.\ Math.\ Phys.\  {\bf 2}, 231 (1998)
  [Int.\ J.\ Theor.\ Phys.\  {\bf 38}, 1113 (1999)]
  [arXiv:hep-th/9711200].

\bibitem{Berenstein:2005aa}
  D.~Berenstein,
  ``Large $N$ BPS states and emergent quantum gravity,''
  JHEP {\bf 0601}, 125 (2006)
  [arXiv:hep-th/0507203].

\bibitem{Berenstein:2005jq}
  D.~Berenstein, D.~H.~Correa and S.~E.~Vazquez,
  ``All loop BMN state energies from matrices,''
  JHEP {\bf 0602}, 048 (2006)
  [arXiv:hep-th/0509015].

\bibitem{Berenstein:2005ek}
  D.~Berenstein and D.~H.~Correa,
  ``Emergent geometry from $q$-deformations of $\N = 4$ super Yang-Mills,''
  JHEP {\bf 0608}, 006 (2006)
  [arXiv:hep-th/0511104].

\bibitem{Berenstein:2006yy}
  D.~Berenstein and R.~Cotta,
  ``Aspects of emergent geometry in the AdS/CFT context,''
  Phys.\ Rev.\ D {\bf 74}, 026006 (2006)
  [arXiv:hep-th/0605220].

\bibitem{Lin:2004nb}
  H.~Lin, O.~Lunin and J.~M.~Maldacena,
  ``Bubbling AdS space and 1/2 BPS geometries,''
  JHEP {\bf 0410}, 025 (2004)
  [arXiv:hep-th/0409174].

\bibitem{Berenstein:2004kk}
  D.~Berenstein,
  ``A toy model for the AdS/CFT correspondence,''
  JHEP {\bf 0407}, 018 (2004)
  [arXiv:hep-th/0403110].


\bibitem{Vazquez:2006id}
  S.~E.~Vazquez,
  ``Reconstructing 1/2 BPS space-time metrics from matrix models and spin chains,''
  arXiv:hep-th/0612014.


\bibitem{Berenstein:2003gb}
  D.~Berenstein, J.~M.~Maldacena and H.~Nastase,
  ``Strings in flat space and pp waves from $\N=4$ Super Yang Mills,''
  AIP Conf.\ Proc.\  {\bf 646}, 3 (2003).

\bibitem{Hofman:2006xt}
  D.~M.~Hofman and J.~M.~Maldacena,
  ``Giant magnons,''
  J.\ Phys.\ A {\bf 39}, 13095 (2006)
  [arXiv:hep-th/0604135].

\bibitem{Vazquez:2006hd}
  S.~E.~Vazquez,
  ``BPS condensates, matrix models and emergent string theory,''
  arXiv:hep-th/0607204.

\bibitem{Dorey:2006dq}
  N.~Dorey,
  ``Magnon bound states and the AdS/CFT correspondence,''
  J.\ Phys.\ A {\bf 39}, 13119 (2006)
  [arXiv:hep-th/0604175].

\bibitem{Chen:2006ge}
  H.~Y.~Chen, N.~Dorey and K.~Okamura,
  ``Dyonic giant magnons,''
  JHEP {\bf 0609}, 024 (2006)
  [arXiv:hep-th/0605155].

\bibitem{Arutyunov:2006gs}
  G.~Arutyunov, S.~Frolov and M.~Zamaklar,
  ``Finite-size effects from giant magnons,''
  arXiv:hep-th/0606126.

\bibitem{Minahan:2006bd}
  J.~A.~Minahan, A.~Tirziu and A.~A.~Tseytlin,
  ``Infinite spin limit of semiclassical string states,''
  JHEP {\bf 0608}, 049 (2006)
  [arXiv:hep-th/0606145].

\bibitem{Spradlin:2006wk}
  M.~Spradlin and A.~Volovich,
  ``Dressing the giant magnon,''
  JHEP {\bf 0610}, 012 (2006)
  [arXiv:hep-th/0607009].

\bibitem{Bobev:2006fg}
  N.~P.~Bobev and R.~C.~Rashkov,
  ``Multispin giant magnons,''
  Phys.\ Rev.\ D {\bf 74}, 046011 (2006)
  [arXiv:hep-th/0607018].

\bibitem{Kruczenski:2006pk}
  M.~Kruczenski, J.~Russo and A.~A.~Tseytlin,
  ``Spiky strings and giant magnons on $S^{5}$,''
  JHEP {\bf 0610}, 002 (2006)
  [arXiv:hep-th/0607044].

\bibitem{Okamura:2006zv}
  K.~Okamura and R.~Suzuki,
  ``A perspective on classical strings from complex sine-Gordon solitons,''
  arXiv:hep-th/0609026.

\bibitem{Kalousios:2006xy}
  C.~Kalousios, M.~Spradlin and A.~Volovich,
  ``Dressing the giant magnon II,''
  arXiv:hep-th/0611033.

\bibitem{Hirano:2006ti}
  S.~Hirano,
  ``Fat magnon,''
  arXiv:hep-th/0610027.

\bibitem{Ryang:2006yq}
  S.~Ryang,
  ``Three-spin giant magnons in $AdS_{5}\times S^{5}$,''
  arXiv:hep-th/0610037.

\bibitem{Chen:2006gq}
  H.~Y.~Chen, N.~Dorey and K.~Okamura,
  ``On the scattering of magnon boundstates,''
  JHEP {\bf 0611}, 035 (2006)
  [arXiv:hep-th/0608047].

\bibitem{Roiban:2006gs}
  R.~Roiban,
  ``Magnon bound-state scattering in gauge and string theory,''
  arXiv:hep-th/0608049.

\bibitem{Chu:2006ae}
  C.~S.~Chu, G.~Georgiou and V.~V.~Khoze,
  ``Magnons, classical strings and beta-deformations,''
  JHEP {\bf 0611}, 093 (2006)
  [arXiv:hep-th/0606220].

\bibitem{Maldacena:2006rv}
  J.~Maldacena and I.~Swanson,
  ``Connecting giant magnons to the pp-wave: An interpolating limit of $AdS_{5}\times S^{5}$,''
  arXiv:hep-th/0612079.

\bibitem{Frolov:2003qc}
  S.~Frolov and A.~A.~Tseytlin,
  ``Multi-spin string solutions in $AdS_{5}\times S^{5}$,''
  Nucl.\ Phys.\ B {\bf 668}, 77 (2003)
  [arXiv:hep-th/0304255].

\bibitem{Frolov:2003xy}
  S.~Frolov and A.~A.~Tseytlin,
  ``Rotating string solutions: AdS/CFT duality in non-supersymmetric sectors,''
  Phys.\ Lett.\ B {\bf 570}, 96 (2003)
  [arXiv:hep-th/0306143].

\bibitem{Chen:2006gp}
  H.~Y.~Chen, N.~Dorey and K.~Okamura,
  ``The asymptotic spectrum of the $\N = 4$ super Yang-Mills spin chain,''
  arXiv:hep-th/0610295.

\bibitem{Lunin:2005jy}
  O.~Lunin and J.~M.~Maldacena,
  ``Deforming field theories with $U(1) \times U(1)$ global symmetry and their gravity duals,''
  JHEP {\bf 0505}, 033 (2005)
  [arXiv:hep-th/0502086].

\bibitem{Leigh:1995ep}
  R.~G.~Leigh and M.~J.~Strassler,
  ``Exactly Marginal Operators And Duality In Four-Dimensional $\N=1$ Supersymmetric Gauge Theory,''
  Nucl.\ Phys.\ B {\bf 447}, 95 (1995)
  [arXiv:hep-th/9503121].

\bibitem{Frolov:2005ty}
  S.~A.~Frolov, R.~Roiban and A.~A.~Tseytlin,
  ``Gauge - string duality for superconformal deformations of $\N = 4$ super Yang-Mills theory,''
  JHEP {\bf 0507}, 045 (2005)
  [arXiv:hep-th/0503192].

\bibitem{Frolov:2005dj}
  S.~Frolov,
  ``Lax pair for strings in Lunin-Maldacena background,''
  JHEP {\bf 0505}, 069 (2005)
  [arXiv:hep-th/0503201].

\bibitem{Beisert:2004hm}
  N.~Beisert, V.~Dippel and M.~Staudacher,
  ``A novel long range spin chain and planar $\N = 4$ super Yang-Mills,''
  JHEP {\bf 0407}, 075 (2004)
  [arXiv:hep-th/0405001].

\bibitem{Arutyunov:2004vx}
  G.~Arutyunov, S.~Frolov and M.~Staudacher,
  ``Bethe ansatz for quantum strings,''
  JHEP {\bf 0410}, 016 (2004)
  [arXiv:hep-th/0406256].

\bibitem{Fischbacher:2004iu}
  T.~Fischbacher, T.~Klose and J.~Plefka,
  ``Planar plane-wave matrix theory at the four loop order: Integrability without BMN scaling,''
  JHEP {\bf 0502}, 039 (2005)
  [arXiv:hep-th/0412331].









\end{thebibliography}
\end{document}